\documentclass[epjCONF]{svjour}
\usepackage{graphics}
\usepackage[varg]{txfonts} 
\usepackage[latin1]{inputenc}
\session-title{MESON2012 - 12th International Workshop on Meson Production, Properties and Interaction}
\newcommand{\scaleFactor}{0.3}
\begin{document}
\title{\large \bf
   Search for $\eta$-mesic Helium with the WASA-at-COSY detector.
\\ }
\author{Wojciech Krzemien\inst{1}\fnmsep\thanks{\email{wojciech.krzemien@if.uj.edu.pl}} \and Pawel Moskal\inst{1}\inst{2} \and Jerzy Smyrski\inst{1} \and Magdalena Skurzok\inst{1}\\ for the WASA-at-COSY collaboration }
\institute{Institute of Physics, Jagiellonian University, Cracow, Poland \and Institut fur Kernphysik (IKP) and J\"ulich Center for Hadron Physics (JCHP), Forschungszentrum J\"ulich, Germany.}
\abstract{
A search for the ${^4\mbox{He}}-\eta$ bound state via  exclusive measurement
of the excitation function for the $dd \rightarrow {^3\mbox{He}} p \pi^-$ reaction, was performed at the Cooler Synchrotron COSY-J\"ulich with the WASA-at-COSY detection system.
The data were taken during a slow acceleration of the beam
from 2.185\,GeV/c to 2.400\,GeV/c crossing the kinematic threshold
for the $\eta$ production in the $dd \rightarrow {^4\mbox{He}}\,\eta$ reaction at 2.336~GeV/c.
The corresponding excess energy in the ${^4\mbox{He}}-\eta$ system varied from -51.4~MeV to 22~MeV.
The shape of the excitation function for the $dd \rightarrow {^3\mbox{He}} p \pi^{-}$ was examined.
No signal of the ${^4\mbox{He}}-\eta$ bound state was observed in the excitation function.
} 
\maketitle
\section{Introduction}
\label{intro}

It is conceivable that neutral mesons such as $\eta, \bar{K},\omega,\eta'$, $J/\Psi$ \cite{hiren,metag,tshushima1,tshushima2,jinr} can form bound states
with atomic nuclei.
In this case the binding is exclusively due to the strong interaction and the bound state - {\em mesic nucleus}
- can be considered as a meson moving in the mean field of the nucleons in the nucleus.
Due to the strong attractive $\eta$-nucleon interaction~\cite{moskalNew,wycech}, the $\eta$-mesic nuclei are ones of the most promising candidates for such states.

The existence of $\eta$-mesic nuclei was postulated in 1986 by Haider and Liu \cite{liu2},
and since then a search for such states  was conducted in many experiments in the past~\cite{lampf,lpi,jinr,gsi,gem,mami,mami2} and is being continued at COSY~\cite{moskalsymposium,jurekhe3,timo,meson08,jurekmeson08}, JINR~\cite{jinr}, J-PARC~\cite{fujiokasymposium} and MAMI~\cite{mami,mami2}.
Many promising indications where reported, however, so far there is no direct experimental confirmation of the existence of mesic nucleus.

A very strong final state interaction (FSI) observed
in the $dd \rightarrow {^4\mbox{He}} \eta$ reaction close to kinematical threshold
and interpreted as possible indication of ${^4\mbox{He}}-\eta$ bound state~\cite{Willis97}
suggests, that ${^4\mbox{He}}-\eta$ system is a good candidate for experimental study of possible binding.
This conclusion is strengthen by the predictions in reference~\cite{wycech}.
However, as it was stated in ~\cite{liu3,haider2}, the theoretical predictions for  width and binding energy of the $\eta$-mesic nuclei are strongly dependent on the not well known  subtreshold $\eta$-nucleon interaction. Therefore, direct measurements which could confirm the existence of the bound state, are mandatory.

\section{Method}
In our experimental studies, we used the deuteron-deuteron collisions at energies around the $\eta$ production threshold for production of the $\eta-{^4\mbox{He}}$ bound state.
We expect, that the decay of such state proceeds via absorption of the $\eta$ meson on one of the nucleons
in the ${^4\mbox{He}}$ nucleus leading to excitation of the  $N^{\star}$(1535)  resonance which subsequently
decays in pion-nucleon pair.
The remaining three nucleons play a role of spectators and they are likely to bind forming ${^3\mbox{He}}$
or ${^3\mbox{H}}$ nucleus.

According to the discussed scheme, there exist four equivalent decay channels
of the $({^4\mbox{He}}-\eta)_{bound}$ state.

In our experiment we concentrated on the ${^3\mbox{He}} p \pi^{-}$ decay mode .
In the case of a similar system, the  ${^{4}_{\Lambda}\mbox{He}}$ hypernucleus, it was observed that in the $\pi^{-}$ decay channel the decay mode $ {^{4}_{\Lambda}\mbox{He}} \rightarrow {^3\mbox{He}} p \pi^{-}$ is dominant~\cite{Fet}. 

\begin{figure}
\begin{center}
      \scalebox{0.25}
         {
              \includegraphics{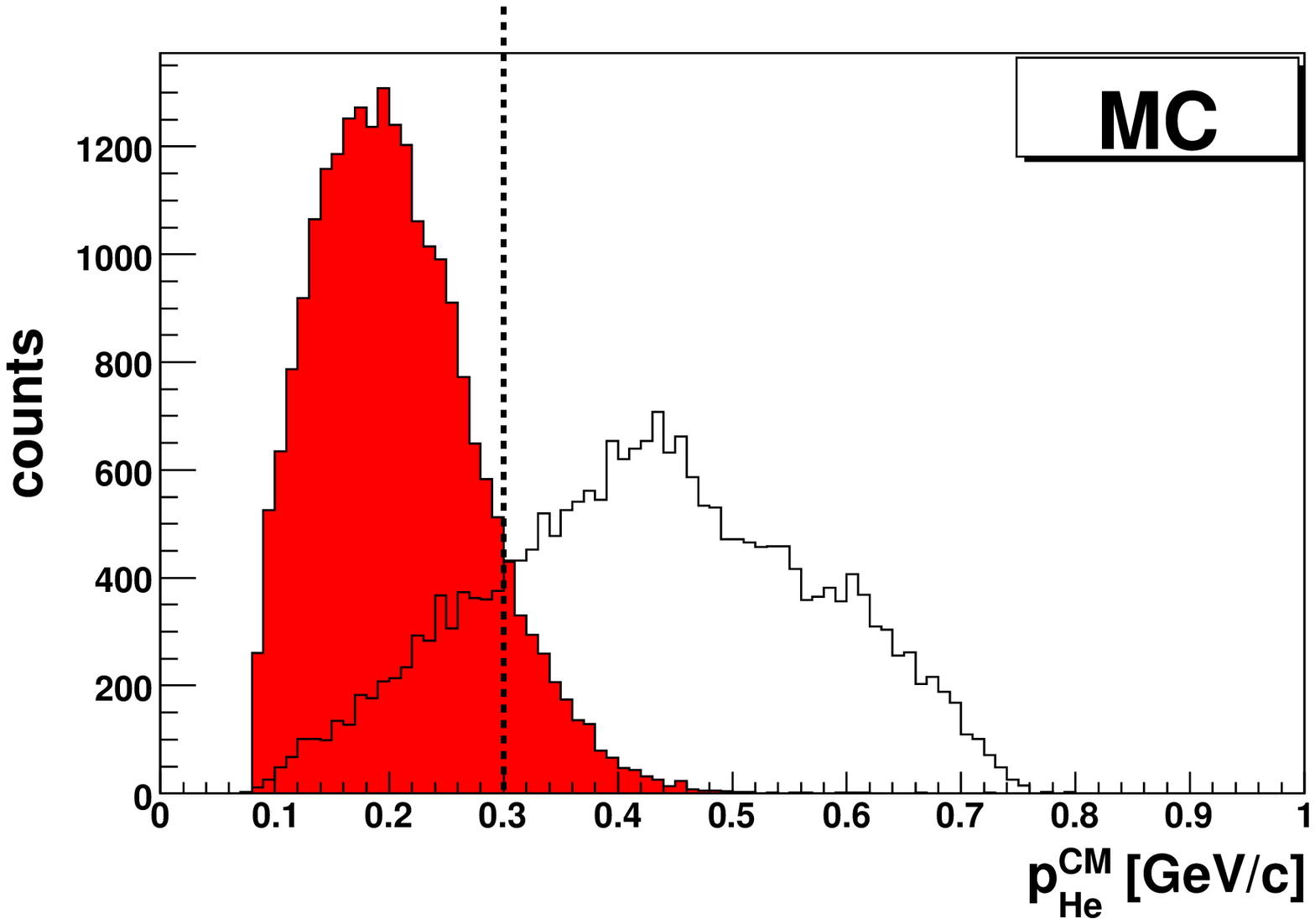}
         }
      \scalebox{0.25}
         {
              \includegraphics{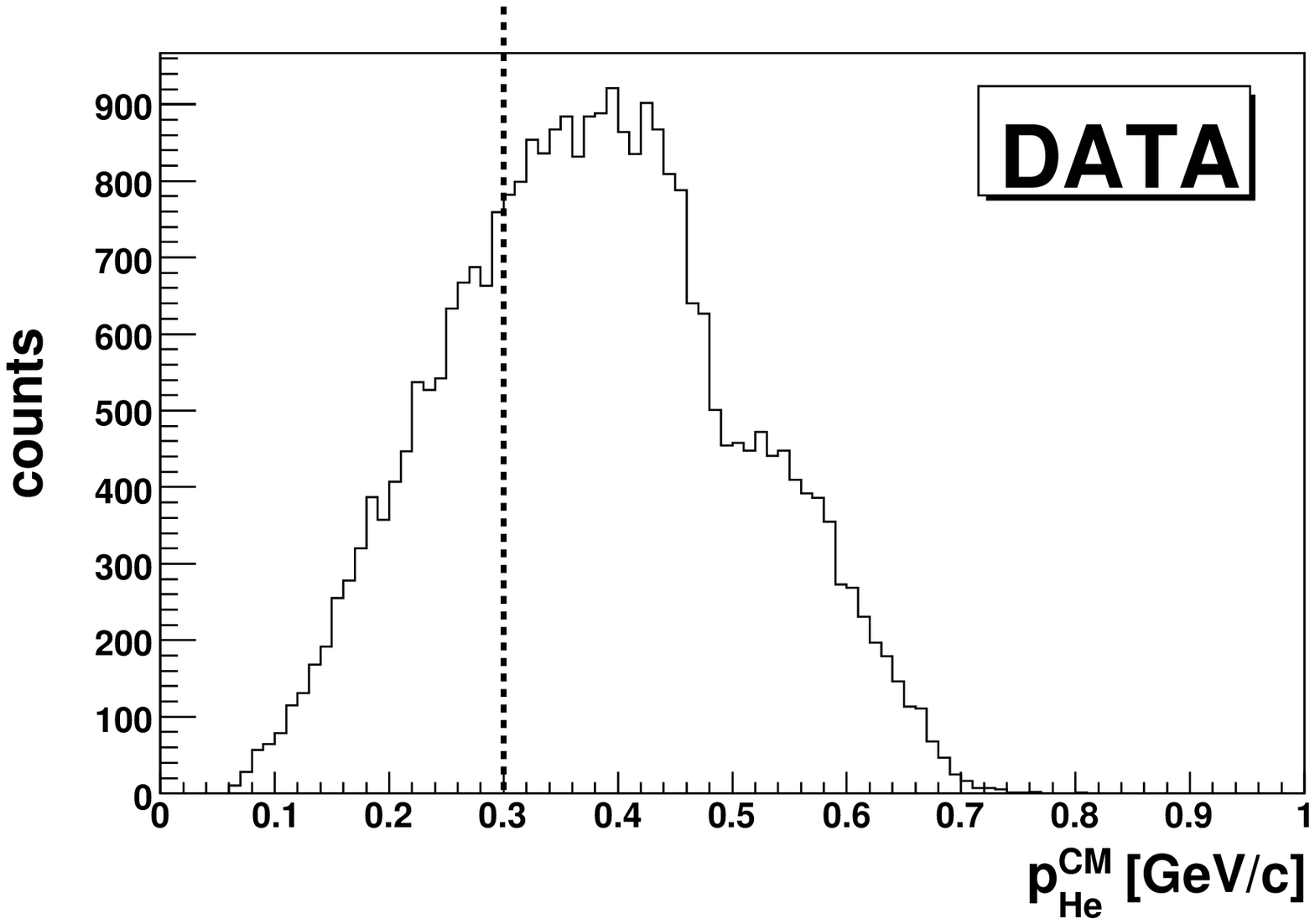}
         }

\caption[Distribution of the ${^3\mbox{He}}$ momentum in CM]{\label{pmomCM_p} (Left plot) Distribution of the ${^3\mbox{He}}$ momentum in 
the CM system simulated for the processes leading to the creation
of the  ${^4\mbox{He}}\eta$ bound state:
$dd \rightarrow (^4{\mbox{He}} \eta)_{bound} \rightarrow {^3{\mbox{He}}}p\pi^{-}$ (red area)
and of the direct $dd \rightarrow {^3{\mbox{He}}} p\pi^{-}$ reaction (black line).
The simulation was done for a momentum of the deuteron beam of 2.307~GeV/c.
The Fermi momentum parametrization was taken from \cite{VH}.
(Rigth plot) Experimental distribution of the ${^3\mbox{He}}$ momentum in the CM system.
In both plots the dashed line demarcates the "signal-poor" and the "signal-rich" regions.
Decrease of the counts at 0.48~GeV/c is due to geometry of the border of the barrel and the end-caps of the Scintillator Barrel detector 
which was used in the $p-\pi^{-}$ identification process.
This region has no relevance  in the next steps of the analysis.
}
\end{center}
\end{figure}

The outgoing $^3\mbox{He}$ nucleus plays the role of a spectator and, therefore,
we expect that its momentum in the CM frame is relatively low and can be described
by the Fermi momentum distribution of nucleons in the $^4\mbox{He}$ nucleus.
This signature  allows to suppress background from reactions leading to the
${^3\mbox{He}} p \pi^{-}$ final state but proceeding without formation of the intermediate
$({^4\mbox{He}}-\eta)_{bound}$ state and, therefore, resulting on the average in much higher
CM momenta of $^3\mbox{He}$ (see Fig.~\ref{pmomCM_p}).

The principle of the present experiment was based on the measurement of the excitation function
of the $dd \rightarrow {^3\mbox{He}} p \pi^{-}$ reaction for energies in the vicinity of the $\eta$ production
threshold and on the selection of events with low ${^3\mbox{He}}$ CM momenta.
In the case of existence of the ${^4\mbox{He}}-\eta$ bound state we expected to observe
a resonance-like structure in the excitation function at the reaction CM energies below the $\eta$ threshold.

\section{Experiment}

In June 2008 we performed a search for the $\eta$-mesic ${^4\mbox{He}}$ by measuring the excitation fun\-ction of the $dd \rightarrow ^3\mbox{He} p\pi^-$  reaction near the $\eta$ meson production threshold using the WASA-at-COSY detector~\cite{jurekmeson08}. During the experimental run the momentum of the deuteron beam was varied continuously within each acceleration cycle
from  2.185~GeV/c to 2.400~GeV/c, crossing the kinematic threshold for the $\eta$ production in the $dd \rightarrow {^4\mbox{He}}\,\eta$ reaction at 2.336~GeV/c.
This range of beam momenta corresponds to the variation of $^4\mbox{He}-\eta$  excess energy  from -51.4~MeV to 22~MeV.

\begin{figure}
\begin{center}
  \scalebox{\scaleFactor}
  {
    \includegraphics{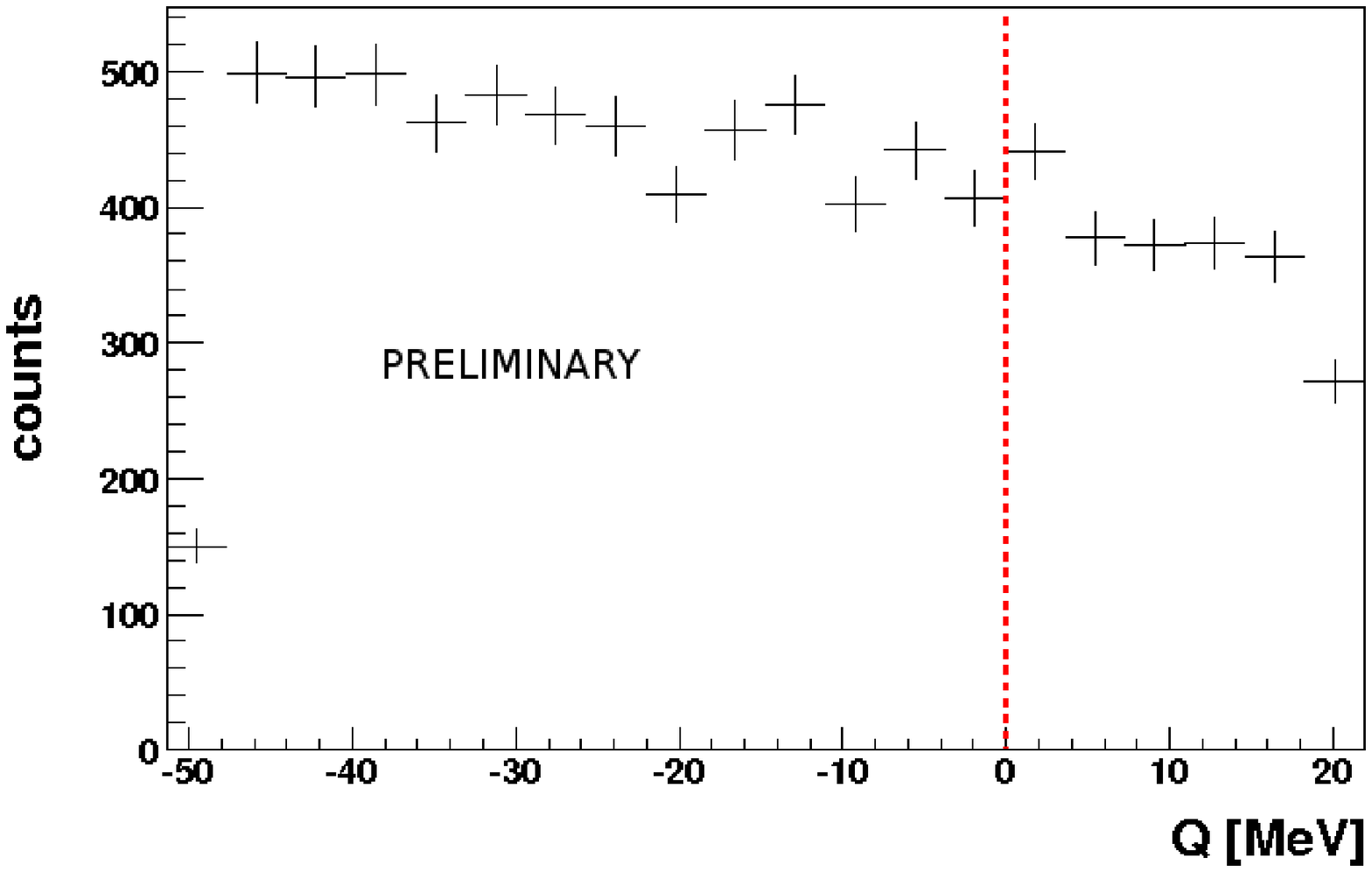}
  }
  \scalebox{\scaleFactor}
  {
    \includegraphics{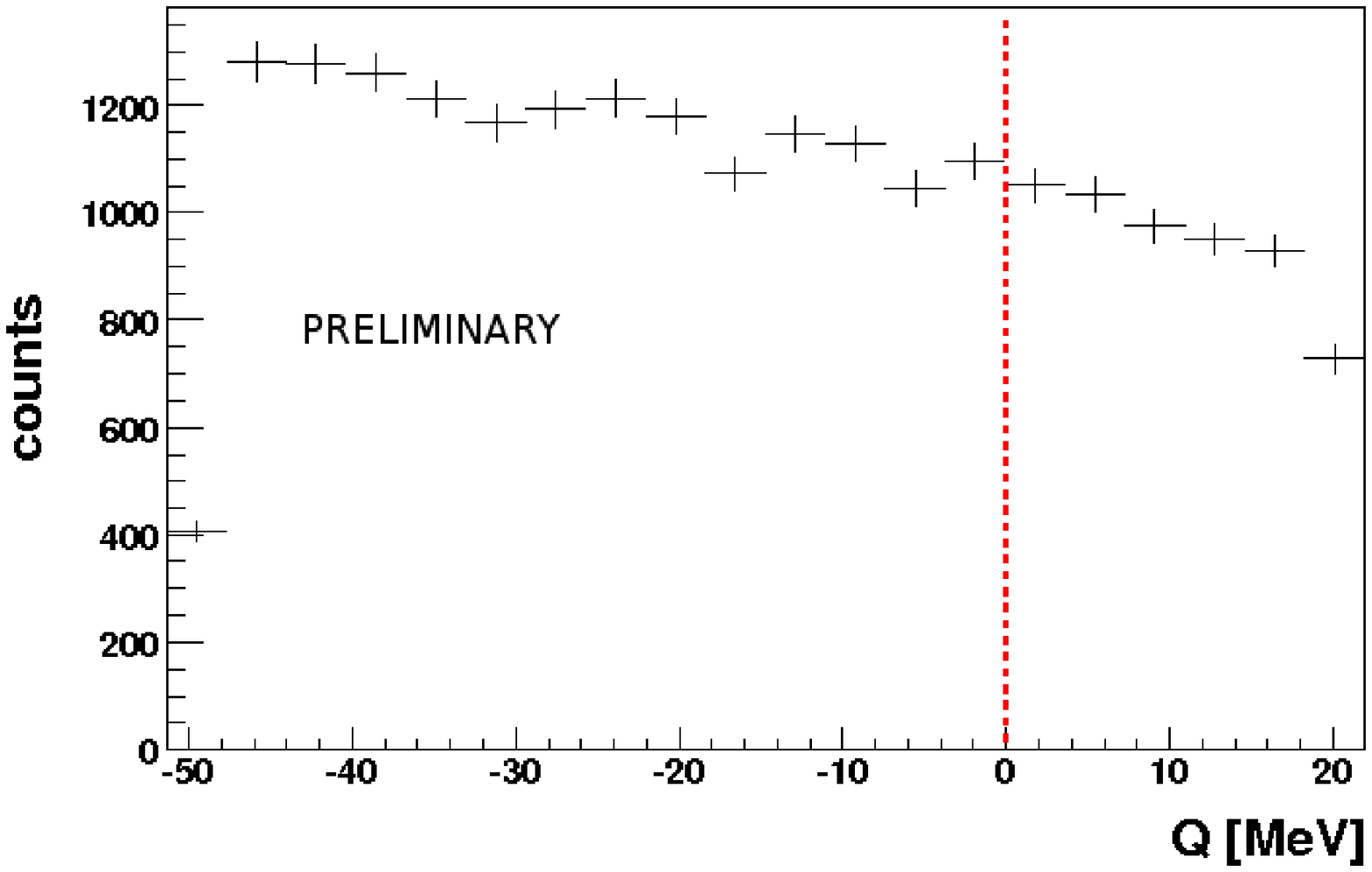}
  }
  \scalebox{\scaleFactor}
  {
    \includegraphics{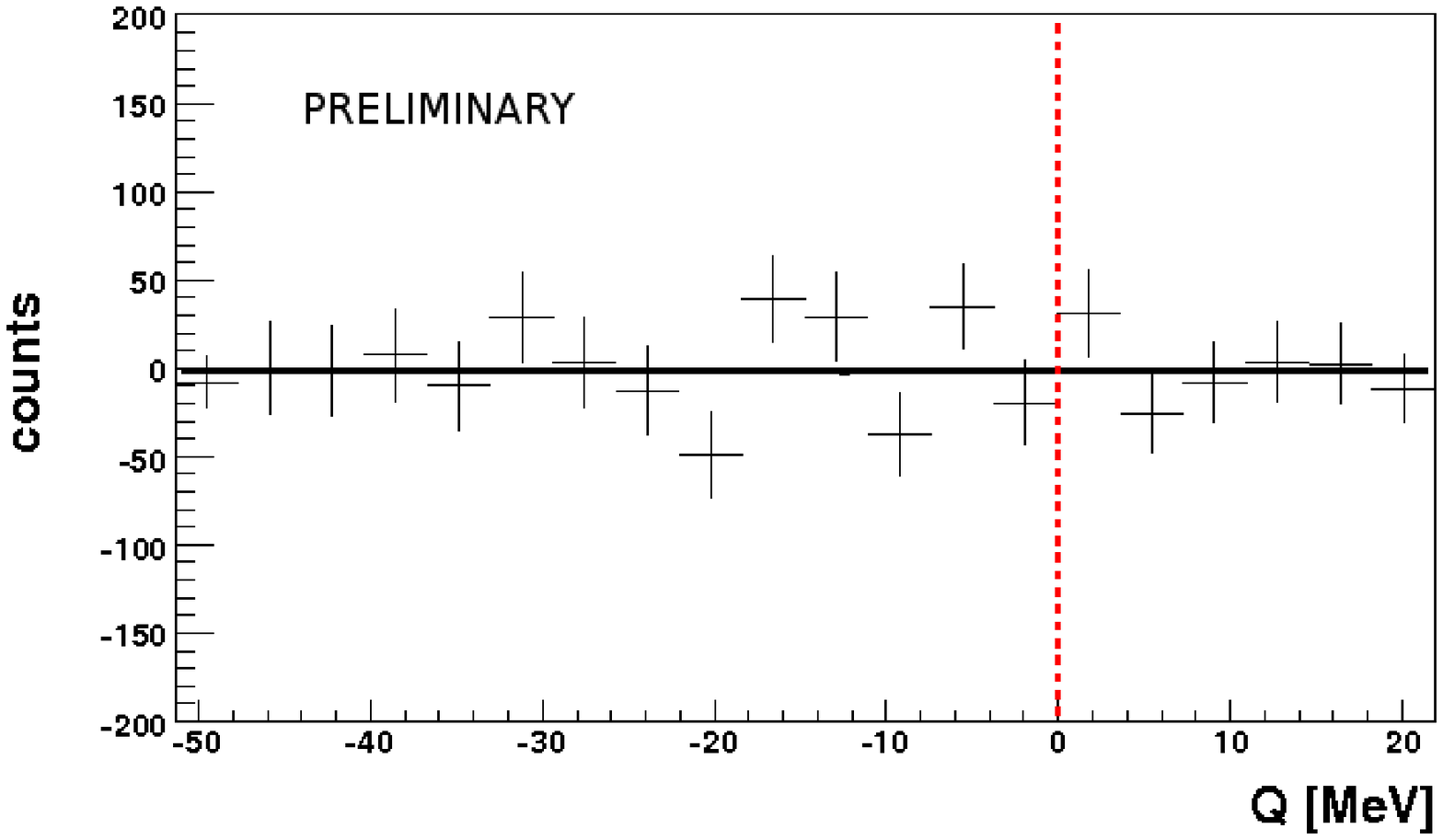}
  }
\caption[Not normalized excitation functions]{\label{hExcitFuncCM_mom_exp_bad_p}Excitation function for the $dd \rightarrow {^3\mbox{He}} p \pi^{-}$ reaction for the "signal-rich" region corresponding to ${^3\mbox{He}}$ momentum below 0.3\,GeV/c (upper left) and the "signal-poor" region with  ${^3\mbox{He}}$ momentum above 0.3\,GeV/c (upper right).
Difference of the excitation functions for the "signal-rich" and "signal-poor" regions after the normalization
to the second bin of Q is shown in the lower panel. The black solid line represents a straight line fit.
The threshold of ${^4\mbox{He}}-\eta$  is marked by the vertical dashed line.}
\end{center}
\end{figure}

We constructed two types of excitation function for the $dd \rightarrow {^3\mbox{He}} p \pi^{-}$ reaction.
They differ in the selection of the events and in the way of normalizing the data points.
The first excitation function uses events from the  "signal-rich" region
corresponding to the ${^3\mbox{He}}$ CM momenta below 0.3\,GeV/c. 
The counts are plotted as a function of the excess energy (Q) as it is shown Fig.~\ref{hExcitFuncCM_mom_exp_bad_p}(top left).
The obtained function is smooth an no clear signal, which could be interpreted as a resonance-like
structure, is visible.
A similar dependence was obtained for events originating from the "signal-poor" region
corresponding to ${^3\mbox{He}}$ CM momenta above 0.3\,GeV/c (see Fig.~\ref{hExcitFuncCM_mom_exp_bad_p}(top rigth)).
We checked also for possible structures in the difference between the discussed functions
for the "signal-rich" and "signal-poor" region.
We multiplied the function for the "signal-poor" region by a factor chosen in such a way,
that the difference of the two functions for the second lowest beam momentum bin is equal to zero.
This difference is presented in Fig.~\ref{hExcitFuncCM_mom_exp_bad_p}(bottom).
The obtained dependence is flat and is consistent with zero. No resonance structure is visible.

In addition, further observables were taken into account in order to reduce the background.

We selected the kinetic energy of protons smaller than 200 MeV  and of pions from the interval (180, 400) MeV.
We applied also a cut on the relative $p-\pi^-$ angle in the CM system in the range of (140$^{\circ}$-180$^{\circ}$).

The absolute value of the integrated luminosity in the experiment was determined 
using the $dd \rightarrow {^3\mbox{He}} n$ reaction and the relative normalization of points of the $dd \rightarrow {^3\mbox{He}} p \pi^-$ excitation
function was based on the quasi-elastic proton-proton scattering~\cite{wkPhD}.

Similarly as in the intermediate stage of the analysis (Fig.~\ref{hExcitFuncCM_mom_exp_bad_p}),
in the final excitation function we observe no structure which could be interpreted
as a resonance originating from the decay of the $\eta$-mesic ${^4\mbox{He}}$.

\section{Outlook}

In November 2010 a new two-week measurement was performed with WASA-at-COSY. We collected  data with approximately 20 times higher statistics. In addition to the $dd \rightarrow {^3\mbox{He}} p \pi^-$ channel we registered also the $dd \rightarrow {^3\mbox{He}} n \pi^{0}$ reaction. The data analysis is undergoing (see ~\cite{Magda2}). 

\section{Support}
This work has been supported by the Polish National Science Center
through grant No. 0320/B/H03/2011/40 and 2011/01/B/ST2/00431,
by the FFE funds of Forschungszetrum J\"ulich, by the European Commission under the 7th Framework Programme through the 'Research Infrastructures' action of the 'Capacities' Programme. Call: FP7-INFRASTRUCTURES-2008-1, Grant Agreement N. 227431 and by the Foundation for Polish Science - MPD program, co-financed by the European Union within the European Regional Development Fund

\end{document}